# Pulse-Driven Reconfiguration of Fractional Polar Topology in Zr-Substituted Barium Titanate

Florian Mayer[1*]

[1]Materials Center Leoben Forschung GmbH, Vordernberger Strasse 12, 8700 Leoben, Austria

*florian.mayer@mcl.at

## Abstract

Polar topological textures in ferroelectrics can host internal structure beyond a single integer topological charge. Here, effective-Hamiltonian molecular-dynamics simulations are used to examine whether such internal fractional topology can be reconfigured by local electric excitation in ordered 12.5% Zr-substituted barium titanate. Chemical doubling along the polar axis stabilizes a coupled nanodomain texture consisting of alternating $Q = -2$ antiskyrmionic and $Q = +4$ skyrmionic slices, in which the local topological charge fragments into six $-1/3$ and six $+2/3$ localized contributions, denoted here as topological quarks, separated by Bloch-point-like singular conversion regions. Picosecond local electric-field pulses applied to selected vortex-core columns drive reconfiguration of the internal dipolar texture of a 2.6 nm nanodomain. Under a binary pulse-mask protocol addressing the six vortex cores, all 64 masks lead, within the chosen low-temperature simulation protocol, to distinct relaxed metastable configurations. The switching calculations are performed in a cryogenic regime, and the programmed states remain stable over at least 1 ns of field-free evolution on the simulation timescale. The resulting configurations are distinguishable both by sector-resolved topological fingerprints and by their real-space polarization fields. These results provide a computational proof of concept that fractional polar topology in a ferroelectric nanodomain can be locally reconfigured by ultrafast electric excitation and used as a multistate configurational degree of freedom in an idealized low-temperature setting.

# I. INTRODUCTION

Topological textures in ferroic materials have emerged as a central theme in contemporary condensed-matter physics [1–5]. Among them, skyrmions and antiskyrmions are particularly attractive because they combine nanoscale dimensions, non-trivial topology, and quasiparticle-like behavior [1,6]. Much of the early development of the field was driven by magnetic systems, where the observation of skyrmion lattices in chiral magnets established a new class of topologically non-trivial spin textures and stimulated broad interest in their dynamics and controllability [5,7–9]. More recently, analogous concepts have been extended to polar materials [10–15], where the relevant order parameter is the electric polarization rather than magnetization. This extension is especially appealing because electric dipoles can, in principle, be addressed directly by electric fields, providing a natural route toward electrically controlled topological states [11]. The experimental discovery of room-temperature polar skyrmions in ferroelectric/paraelectric oxide superlattices [13], together with earlier observations of polar vortices [14], established that ferroelectric oxides can host stable topological polarization textures under suitable electrostatic, elastic, and interfacial boundary conditions. Subsequent work revealed a broad range of polar topologies, including skyrmion bubbles, merons, skyrmion bags, field-driven topological transformations, and emergent topological polarization textures in chemically complex relaxor ferroelectrics [11,12,16–19]. Recent work has also demonstrated deterministic electrical generation of a single electric skyrmion bubble in a ferroelectric nanostructure, further underscoring the interest in electrically driven polar topological transformations [20]. These developments have established polar topology as a distinct arena for studying nontrivial real-space structures in ferroic matter.

Beyond the existence of polar topological textures themselves, a central question concerns the richness of their internal structure and its degree of controllability. In many cases, topological states are classified primarily by a single integer charge, such as the skyrmion number. However, recent work has shown that polar nanodomains can host a much more intricate internal topology. In $PbTiO_3$, skyrmion bubbles were predicted to emerge from columnar ferroelectric nanodomains embedded in a matrix of opposite polarization [21]. In rhombohedral $BaTiO_3$ (BT) [22], even more unusual textures were identified: stable polar antiskyrmions with total topological charge $Q=-2$ [23]. These antiskyrmions differ from conventional $Q=\pm1$ skyrmions not only by the magnitude and sign of their topological charge, but also by the way this charge is distributed within the texture. Rather than being spatially uniform, the total charge fragments into six localized contributions of approximately $-1/3$, referred to as topological quarks [23,24]. For larger nanodomain diameters, each $-1/3$ quark can further split into two $-1/6$ pre-quarks [24], revealing a hierarchy of fractional topological structure within a single ferroelectric nanodomain.

This picture has been further enriched by the prediction of fractional Skyrme lines [25] and by the discovery of switchable skyrmion–antiskyrmion tubes in rhombohedral BT and related materials [26]. Together, these studies [23–26] show that BT, despite being an archetypal ferroelectric, supports a remarkably rich topological phase space. External-field control of nonmagnetic antiskyrmions has already been demonstrated at the mesoscale within a Ginzburg–Landau–Devonshire framework through field-driven creation, annihilation, and transport [27]. A particularly important next question is therefore whether the internal fractional topology of a single such nanodomain can be manipulated in a controlled way by local electric excitation.

A particularly natural route toward such control is chemical design. BT is an archetypal ferroelectric oxide whose phase behavior and local-mode energetics are highly sensitive to B-site substitution. In $Ba(Zr,Ti)O_3$ (BZT) [28–35], Zr substitution modifies the local anisotropy, short-range interactions, and stability of polar nanostructures while preserving the perovskite framework. Across composition, BZT exhibits a rich evolution from BT-like ferroelectric behavior at low Zr content toward increasingly disordered and relaxor-like behavior at higher Zr concentrations [31,32]. In the dilute-to-low-substitution regime, the Ti-rich ferroelectric matrix remains capable of supporting coherent polar textures, while the Zr ions locally perturb the polarization landscape and can act as a chemical handle for stabilizing structures that are absent or less favorable in pure BT [28–30]. This makes low-Zr BZT a useful model system for asking how controlled chemical modulation affects polar topology.

Recent work on Zr-substituted BT demonstrated that substitutional ordering can stabilize new polar topological textures [36]. In the ordered 12.5% structure considered here, one Zr ion per eight B-site positions produces a chemical periodicity doubled along the rhombohedral [111] direction. This structure is used as an idealized chemically modulated model system, not as a representation of generic randomly substituted bulk BZT. It provides a controlled limit in which the effect of a well-defined B-site modulation on the local anisotropy landscape and nanodomain topology can be isolated. The doubled period creates two inequivalent (111) slices within the same nanodomain repeat unit. Previous calculations [36] showed that these slices host a coupled Q=−2/Q=+4 texture: one slice contains a Q=−2 antiskyrmionic state with six localized approximately −1/3 topological charge contributions, whereas the other contains a Q=+4 skyrmionic state with six approximately +2/3 contributions. The two slices share a common six-vortex scaffold, while Bloch-point-like [37–39], lattice-regularized conversion regions mediate the integer offset between their slice-resolved topological charges.

The static Q=−2/Q=+4 texture raises the central dynamical question addressed here: can its internal fractional topology be reconfigured by local electric excitation? The six vortex-core columns provide natural control coordinates for this purpose. They are spatially localized, carry pronounced in-plane dipoles, and belong to the common six-vortex scaffold that connects the two inequivalent (111) slices. A perturbation of one core is therefore not expected to remain purely local. Instead, it can couple to the surrounding scaffold and to the Bloch-point-like conversion regions that mediate the offset between the Q=−2 and Q=+4 sectors. Reversing or strongly deflecting a core dipole may therefore provide a local trigger for redistributing fractional topological charge within the coupled nanodomain.

This idea motivates the excitation protocol used in the present work. Rather than applying a spatially homogeneous field to the entire nanodomain, localized picosecond electric-field pulses are applied to selected vortex-core columns. For each addressed core, the field is oriented parallel to the corresponding vortex-core dipole in the reference state. In the (111) projection, this direction is tangential with respect to the nanodomain center and therefore couples directly to the local core-reversal coordinate. Binary pulse masks then select different subsets of the six cores. These masks should be understood as controlled excitation pathways through a continuous polarization-field landscape, not as an assumption that the underlying metastable state space is intrinsically binary.

If such local excitation can reproducibly access distinct field-free configurations, fractional polar topology would become more than a static structural feature: it would constitute a dynamically programmable configurational degree of freedom of a ferroelectric nanodomain. This possibility is tested here using effective-Hamiltonian molecular dynamics in ordered 12.5% Zr-substituted BT. Starting from the previously established [36] coupled Q=−2/Q=+4 texture, selected vortex-core columns are excited in a cryogenic low-temperature regime using the local pulse protocol described above. The $2^6$=64 pulse masks are used to sample metastable configurations accessible to the compact 2.6 nm nanodomain. The resulting field-free relaxed states are analyzed in terms of their stability on the simulation timescale, sector-resolved topological charge, inter-slice charge transfer, and real-space polarization structure. A constrained energetic analysis is also used to verify that reversal of a vortex-core column connects locally stable configurations separated by a finite barrier. The results show that the chosen pulse protocol generates 64 distinguishable metastable configurations, stable for at least 1 ns of field-free evolution at 10 K. These findings provide a computational proof of concept that the internal fractional topology of a ferroelectric nanodomain can be locally reconfigured by ultrafast electric excitation in an idealized low-temperature setting.

# II. COMPUTATIONAL METHODOLOGY

All computational experiments were performed using molecular-dynamics (MD) simulations based on first-principles-parameterized effective Hamiltonians (EH) [40–45]. Effective Hamiltonians provide a computationally efficient framework while retaining a quantitative description of the relevant atomistic degrees of freedom and their couplings [29,46–50]. The parametrizations employed here were developed by Mayer et al. [30,51], including anharmonic couplings for pure BT and an improved treatment of substitution effects in Zr-substituted BT. Details of the Hamiltonian for pure BT are given in Ref. [51], whereas the Zr-substituted extension is described in Ref. [30]. An important advantage of these parametrizations is their accurate reproduction of phase-transition temperatures in both pure and substituted BT [30,51], which is essential for a physically meaningful description of the topological textures considered here.

Within the EH framework, the microscopic order parameter is the B-site-centered local-mode vector $\boldsymbol{u}(\boldsymbol{R})$, where $\boldsymbol{R}$ labels the five-atom perovskite unit cell within the simulation supercell [46,47]. This quantity is directly associated with the local dipole moment [46] and, equivalently, with the local polarization field. For topological analysis [36], the normalized order-parameter field is used,

$$\boldsymbol{n}(\boldsymbol{R}) = \frac{\boldsymbol{u}(\boldsymbol{R})}{|\boldsymbol{u}(\boldsymbol{R})|} \tag{1}$$

or equivalently $\boldsymbol{n}(\boldsymbol{R}) = \boldsymbol{p}(\boldsymbol{R})/|\,\boldsymbol{p}(\boldsymbol{R})\,|$, where $\boldsymbol{p}(\boldsymbol{R})$ denotes the local polarization. This normalization is required because the topological charge is defined for a unit-vector field.

The applied electric field was coupled to the local modes through the dipolar term of the effective Hamiltonian. For a spatially and temporally varying field $\boldsymbol{E}(\boldsymbol{R}, t)$, the field contribution was written as

$$H_E(t) = -\sum_{\boldsymbol{R}} Z^*(\sigma_{\boldsymbol{R}})\boldsymbol{E}(\boldsymbol{R}, t) \cdot \boldsymbol{u}(\boldsymbol{R}) \tag{2}$$

where $\sigma_{\boldsymbol{R}}$ denotes the B-site species of cell $\boldsymbol{R}$. The Born effective charge $Z^*(\sigma_{\boldsymbol{R}})$ is therefore species dependent, so that Ti- and Zr-centered local modes couple to the applied field with different strengths. The corresponding field force entering the molecular-dynamics equations is

$$\boldsymbol{F}_E(\boldsymbol{R}, t) = -\frac{\partial H_E}{\partial \boldsymbol{u}(\boldsymbol{R})} = Z^*(\sigma_{\boldsymbol{R}})\boldsymbol{E}(\boldsymbol{R}, t) \tag{3}$$

The values of $Z^*$ are those of the substituted effective-Hamiltonian parameterization and were obtained from cubic $BaTiO_3$ and $BaZrO_3$ reference calculations [30]. They are treated as fixed B-site-species parameters, consistently with the construction of the local-mode Hamiltonian [30]. Environment-dependent corrections to $Z^*$ are not included and would primarily renormalize the quantitative field scale of the pulse coupling. The complete form of the substituted Hamiltonian and the field-coupling convention are summarized in the Supplemental Material [52].

All effective Hamiltonian molecular-dynamics (EHMD) simulations were carried out using a modified version of the feram code [46,47,49]. The simulations were performed in the canonical ensemble, with temperature control implemented by a velocity-scaling algorithm, validated against Nosé–Poincaré [53] dynamics. A time step of 2 fs was used throughout. Unless noted otherwise, each relaxation trajectory had a total duration of 1 ns, consisting of 800 ps equilibration followed by 200 ps production. Periodic boundary conditions were applied in all three spatial directions.

For the pulse-switching study presented here, the simulation cell was fixed at 32×32×32 unit cells. The pulse-excitation simulations were built on the ordered 12.5% Zr-substituted BT superlattice discussed in the following section. First, the superlattice was relaxed for 1 ns at 10 K. A cylindrical nanodomain with polarization reversed with respect to the surrounding matrix along [111] was then introduced and relaxed for another 1 ns, yielding the metastable 2.6 nm reference texture [36] used as the common initial state for all switching calculations. The temperature of 10 K was chosen as a compromise between numerical stability and dynamical accessibility. It lies well below the thermal-collapse temperature of the 2.6 nm reference nanodomain, while still allowing sufficient local relaxation during the pulse-driven evolution (the corresponding thermal-stability analysis is presented in the Supplemental Material [52]). Starting from this relaxed reference state, local electric-field pulses were applied dynamically during the MD trajectory according to the excitation protocol defined in Sec. III.A. The implemented framework allows the selection of arbitrary pulse masks over the six labeled vortex cores and was used to generate the pulse-driven configurations analyzed here. After the pulse sequence, the simulation was continued for 1 ns without the writing field in order to assess post-pulse metastability. The final relaxed dipolar configuration was obtained by time averaging over the last 200 ps of the field-free trajectory.

For the analysis of the topological textures [24,36], the simulation cell was rotated such that the nanodomain axis defines $z'\parallel[111]$, while $x'$ and $y'$ span the corresponding (111) plane. Topological quantities were evaluated on cross sections normal to the nanodomain axis, i.e. on (111) slices. In the ordered 12.5% superlattice, the chemical doubling along [111] implies that two inequivalent slices must be distinguished within the doubled unit, although the underlying geometrical stacking of (111) planes remains unchanged [36]. Slice-resolved local topological charge densities and plane-integrated charges were evaluated in this rotated representation.

To characterize topology, the Pontryagin density was used,

$$Q = \frac{1}{4\pi} \int_{\Omega} \boldsymbol{n} \cdot \left(\partial_x \boldsymbol{n} \times \partial_y \boldsymbol{n}\right) d^2 r \tag{4}$$

which quantifies the topological charge of a continuous unit-vector field $\boldsymbol{n}(\boldsymbol{r})$ defined on a two-dimensional domain and mapping to the target sphere $S^2$, corresponding to the homotopy group $\pi_2(S^2) = \mathbb{Z}$.

For the actual numerical evaluation of the topological charge on the lattice, the discretized formulation of Berg and Lüscher [54] was employed, as adapted to topological switching problems by Heo et al. [55]. In this approach,

$$Q = \frac{1}{4\pi} \sum_{l} A_l \tag{5}$$

where the sum runs over spherical triangles of signed area $A_l$ spanned by three neighboring normalized vectors $\boldsymbol{n}_i$, $\boldsymbol{n}_j$, and $\boldsymbol{n}_k$. The signed area is computed from

$$\cos\left(\frac{A_l}{2}\right) = \frac{1 + \boldsymbol{n}_i \cdot \boldsymbol{n}_j + \boldsymbol{n}_j \cdot \boldsymbol{n}_k + \boldsymbol{n}_k \cdot \boldsymbol{n}_i}{\sqrt{2(1 + \boldsymbol{n}_i \cdot \boldsymbol{n}_j)(1 + \boldsymbol{n}_j \cdot \boldsymbol{n}_k)(1 + \boldsymbol{n}_k \cdot \boldsymbol{n}_i)}} \tag{6}$$

with the sign determined by

$$sign(A_l) = sign\left[\boldsymbol{n}_i \cdot \left(\boldsymbol{n}_j \times \boldsymbol{n}_k\right)\right].$$

(7)

In the present case, the spin vectors used in magnetic applications are replaced by the normalized local-mode vectors obtained from the EH simulations. This formulation was used to compute both local topological charge densities and plane-integrated topological charges on the (111) slices discussed throughout the manuscript.

# III. RESULTS

This section examines how ultrafast local electric-field pulses reconfigure the coupled fractional topological texture in ordered 12.5% Zr-substituted BT. The analysis proceeds in four steps. First, the Q=−2/Q=+4 reference state and the vortex-core-based local excitation protocol are specified. Second, representative switching events are used to identify the microscopic mechanism of pulse-induced topological reconfiguration. Third, the full $2^6$=64 pulse-mask set is analyzed in terms of its local and slice-resolved topological fingerprints. Finally, the distinguishability of the programmed states is examined directly in the real-space polarization field, establishing that the pulse-addressed state space is reflected in the underlying dipolar structure as well as in its topological representation.

## A. Excitation protocol for topological reconfiguration

The pulse protocol is defined with respect to the ordered 12.5% Zr-substituted BT reference nanodomain introduced in Ref. [36] and discussed in the Introduction. In this structure, one Ti ion per eight B-site positions is replaced by Zr, producing a chemically doubled period along the rhombohedral [111] direction. The simulation cell is analyzed in a rotated coordinate system with $z' \parallel [111]$, while $x'$ and $y'$ span the corresponding (111) plane. The chemical doubling creates two inequivalent (111) slices within the doubled unit, denoted here as $z_1'$ and $z_2'$, which provide the natural basis for the slice-resolved topological analysis.

After relaxation, the cylindrical nanodomain forms the coupled Q=−2/Q=+4 texture shown in Fig. 1. The lower slice $z_1'$ contains a Q=−2 antiskyrmionic texture with six localized approximately −1/3 topological charge contributions [Fig. 1(a)]. The upper slice $z_2'$ contains a Q=+4 skyrmionic texture with six approximately +2/3 localized contributions [Fig. 1(b)]. The two slices share the same six-vortex scaffold but differ in their core-level polarization structure and in their slice-resolved topological charge. The transition between the two local topological sectors occurs through Bloch-point-like, lattice-regularized conversion regions at corresponding in-plane positions [36].

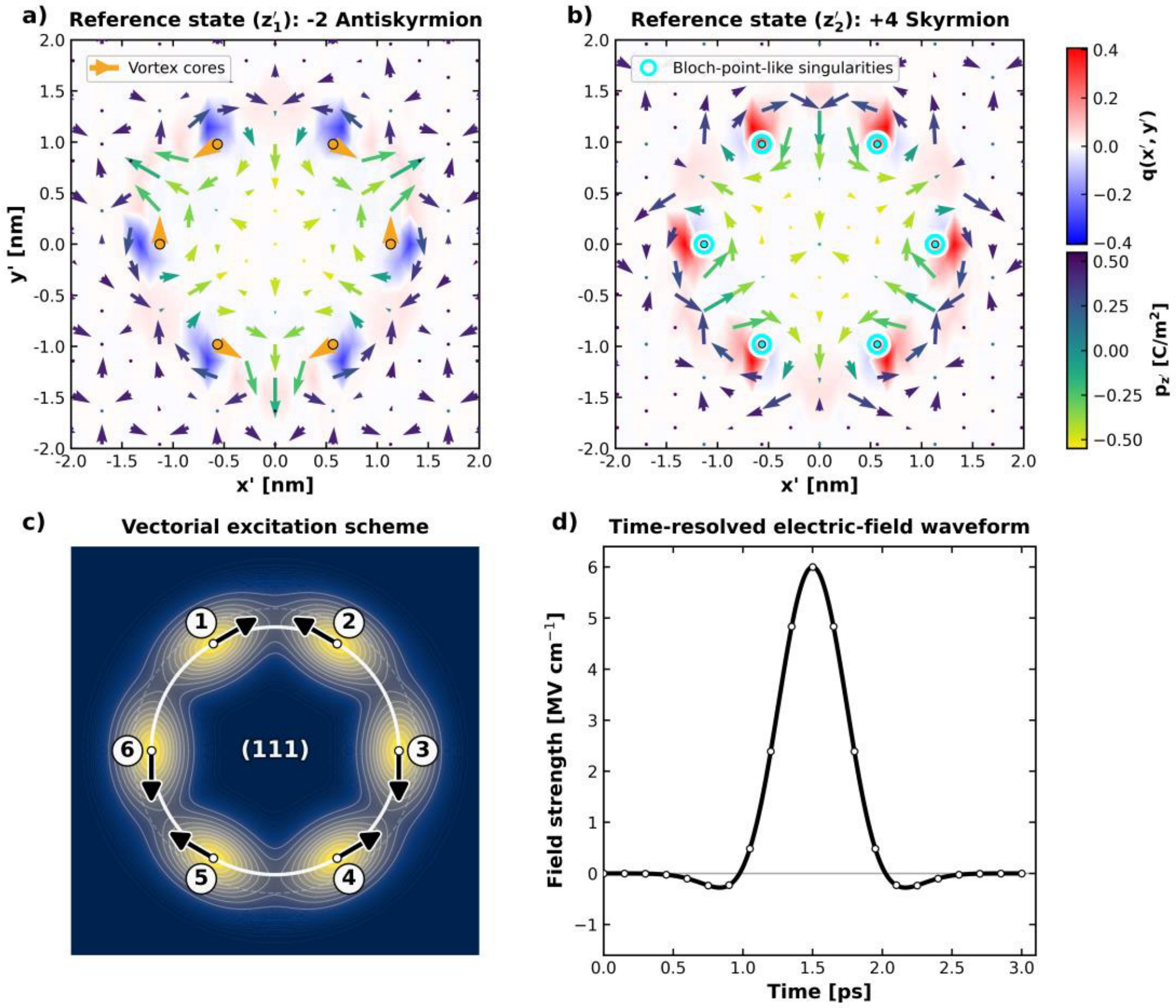

*Figure 1. Reference Q=−2/Q=+4 nanodomain texture and pulse-addressing protocol. (a) (111)-plane projection of the lower half of the doubled unit cell ($z'_1$), showing the Q=−2 antiskyrmionic slice with six localized −1/3 topological quarks. Orange vectors indicate the vortex-core dipoles. (b) Corresponding projection of the upper half ($z'_2$), showing the Q=+4 skyrmionic slice with six localized +2/3 topological quarks. Circled regions indicate Bloch-point-like singular conversion motifs. (c) Labeling of the six vortex cores used to define the binary pulse mask, together with a schematic of the tangential local excitation field. (d) Temporal profile of the electric-field pulse used for pulse-induced reconfiguration. In (a) and (b), polarization vectors are overlaid on the local topological charge density and colored by the out-of-plane polarization component.*

The six vortex-core regions of the Q=−2 slice provide the local excitation centers. They are spatially localized near the nanodomain perimeter, possess pronounced tangential in-plane core dipoles, and are coupled to the corresponding Bloch-point like conversion regions in the neighboring Q=+4 slice. To verify that these core dipoles are not merely passive structural features, a constrained potential-energy-surface analysis was performed for a selected vortex-core column. In this calculation, the local-mode vector of the addressed core column was constrained on a three-dimensional grid, while all remaining local-mode degrees of freedom were relaxed for each imposed core configuration. Two locally stable minima, denoted M1 and M2, were identified on this constrained landscape. A discrete minimum-barrier path search through the sampled grid was then used to connect the two minima, and the corresponding relative total energy of the 32×32×32 supercell is shown in Fig. 2. The two minima differ mainly by the orientation of the tangential core component, and the connecting path contains a finite barrier. This demonstrates that the addressed vortex-core column represents a switchable local coordinate embedded in the coupled six-vortex scaffold. Further details of the constrained landscape, path search, and relaxed polarization configurations are provided in the Supplemental Material [52].

This observation motivates the pulse geometry used below. The local electric field is oriented along the tangential direction of the addressed vortex-core dipole, so that it couples directly to the core-reversal coordinate identified by the constrained energy profile. The pulse is therefore not intended to create a new topological texture from an otherwise unique local minimum. Instead, it drives an existing local degree of freedom across a finite barrier, after which the surrounding polarization field relaxes within the coupled Q=−2/Q=+4 nanodomain environment. Because the core column is coupled to the conversion regions connecting the two inequivalent slices, such a local reversal can redistribute the fractional topological charge rather than only reverse a local dipole.

For the relaxed reference state, the six cores are labeled as shown in Fig. 1(c). A binary mask $m_\alpha$, with α=1,…,6, specifies whether a pulse is applied to core $\alpha$. The mask labels define excitation pathways only. They do not imply that the internal state space is intrinsically binary, since the local modes form a continuous three-dimensional vector field.

The pulse-switching calculations are performed for the relaxed reference nanodomain, whose final diameter is approximately 2.6 nm after zero-field relaxation. This compact texture was selected because it supports the coupled Q=−2/Q=+4 six-vortex topology while remaining metastable in the low-temperature regime. The coupled topology is not restricted to this diameter. Related relaxed nanodomains are also obtained from other initial domain sizes, and larger domains generally show higher thermal stability. The corresponding size-dependent stability analysis is given in the Supplemental Material [52]. The switching simulations are performed at 10 K, well below the collapse temperature of the 2.6 nm reference nanodomain.

The excitation field was localized in the transverse (x′,y′) plane around the selected vortex cores and extended uniformly along z′, so that the perturbation acted on vortex-core columns rather than on isolated (111) slices. For a given binary pulse mask, the field acting on local-mode site $\boldsymbol{R}$ was implemented in the modified feram code as

$$\boldsymbol{E}(\boldsymbol{R},t) = \sum_{\alpha=1}^{6} m_\alpha\, E_0 \exp\left(-\frac{|\boldsymbol{\rho}(\boldsymbol{R}) - \boldsymbol{\rho}_\alpha^c|^2}{\sigma_E^2}\right) g(t)\hat{\boldsymbol{e}}_\alpha, \tag{8}$$

with

$$g(t) = \exp\left(-\frac{(t-t_0)^2}{\Delta^2}\right)\cos\big(\omega(t-t_0)\big) \tag{9}$$

Here $\boldsymbol{\rho}_\alpha^c$ is the transverse position of vortex-core column $\alpha$, $\boldsymbol{\rho}(\boldsymbol{R})$ is the transverse coordinate of local-mode site $\boldsymbol{R}$, and $\hat{\boldsymbol{e}}_\alpha$ is the tangential direction of the local core dipole in the reference state. The Gaussian envelope is independent of $z'$, producing the localized columnar excitation profile sketched in Fig. 1(c). The width parameter was chosen such that the field amplitude is reduced to approximately 20% of its maximum value at a transverse distance of 4 Å from the addressed core, giving $\sigma_E = 4/\sqrt{\ln 5} \simeq 3.15$ Å for the convention used in Eq. (8). The temporal dependence $g(t)$ was chosen as a Gaussian-envelope carrier pulse of the form used in ultrafast effective-Hamiltonian simulations [56] of ferroelectric switching and is shown in Fig. 1(d). Unless stated otherwise, the pulse parameters are $E_0$ = 6 MV cm$^{-1}$, $\Delta$ = 0.45 ps, and $\omega$ = 3 ps$^{-1}$. Pulse-duration tests for the single-core mask 100000 are summarized in the Supplemental Material [52] and show the expected amplitude-duration tradeoff for persistent core reversal. For masks containing more than one addressed core, the selected local fields were applied simultaneously by linear superposition.

All pulse-excitation calculations therefore start from the same relaxed 2.6 nm Q=−2/Q=+4 reference nanodomain. After field removal, the system is evolved at 10 K in zero field, and the final state is obtained from the field-free time-averaged configuration as described in Sec. II. This protocol provides a systematic way to test whether local perturbations of selected vortex-core columns relax into distinct metastable configurations.

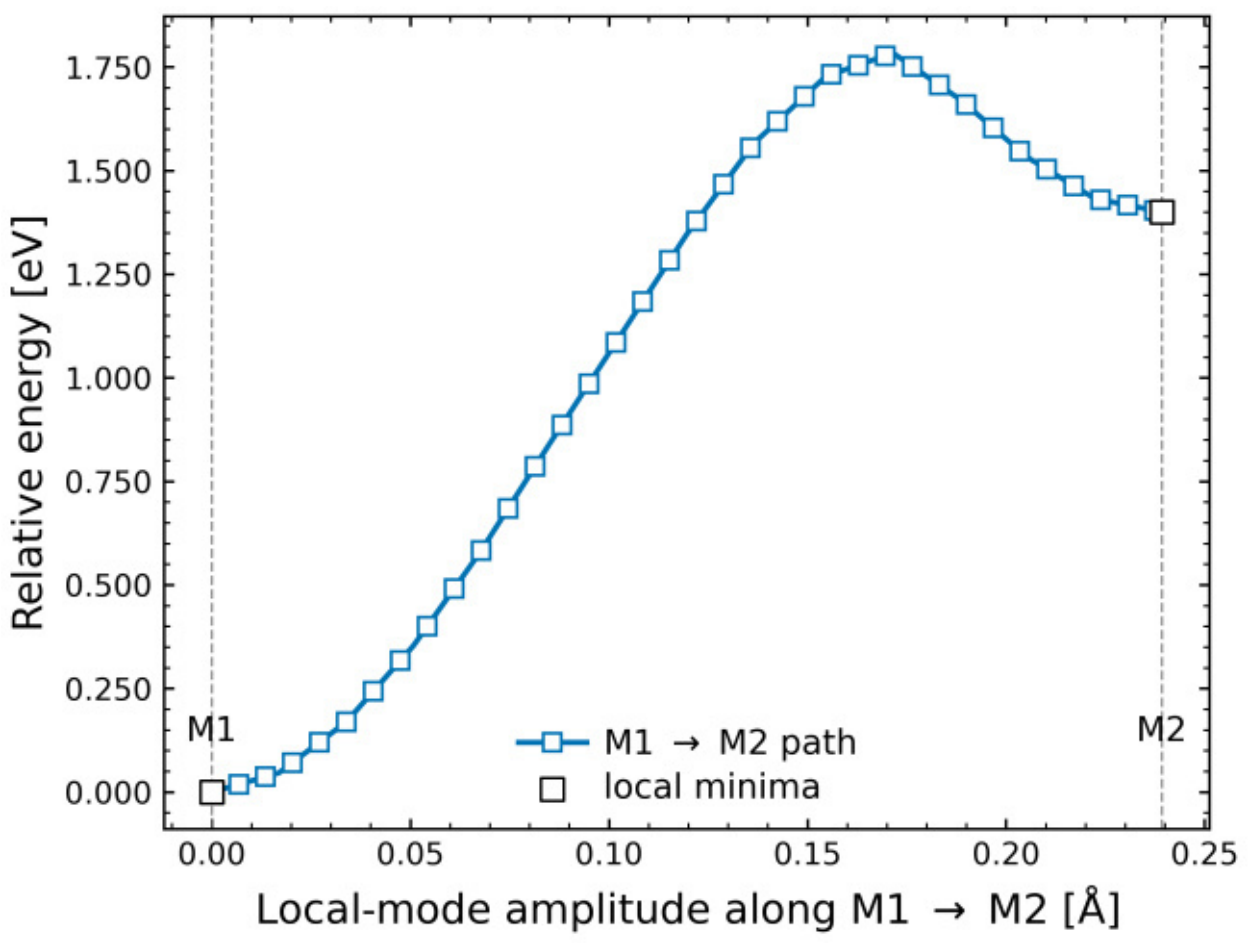

*Figure 2. Constrained energy profile for vortex-core-column reversal. Relative total energy of the 32x32x32 simulation supercell along a constrained M1 → M2 path for reversal of the tangential local-mode component of a selected vortex-core column. The endpoints M1 and M2 correspond to locally stable configurations with approximately opposite tangential core orientations, while the maximum defines a finite switching barrier of about 1.8 eV relative to M1. This barrier motivates the use of tangential picosecond electric-field pulses to drive the core-reversal coordinate.*

## B. Pulse-induced reconfiguration of the coupled topological texture

This section examines how local picosecond excitation of selected vortex cores reconfigures the coupled Q=−2/Q=+4 nanodomain texture. The analysis is based on the real-space polarization fields in the two inequivalent (111) slices, together with the corresponding local topological charge densities. The aim is to identify the main switching pathway observed in the simulations and to establish how local pulse excitation modifies the fractional topological structure of the nanodomain.

The switching process is most naturally understood as follows. The local pulse acts on a vortex-core column that already constitutes a switchable degree of freedom within the coupled nanodomain. As shown by the constrained potential-energy-surface analysis in Fig. 2 and in the Supplemental Material [52], reversal of a single vortex-core column connects two locally stable configurations separated by a finite barrier. The role of the pulse is therefore to drive the selected core coordinate across this barrier. The resulting topological state is then not determined by the local core alone, but by relaxation of the full six-vortex scaffold, including the electrostatic, short-range, elastic, and topological couplings between the two inequivalent slices.

The core-level switching dynamics is illustrated in Fig. 3 for the single-core pulse mask 100000. The local field is applied along the tangential direction of vortex core 1, and the local-mode response is resolved into tangential, radial, and out-of-plane components. During the pulse, the tangential component is driven through zero and changes sign, demonstrating that the field pushes the core across the reversal coordinate identified in Fig. 2. After the pulse has decayed, the tangential component remains reversed during the subsequent field-free evolution, showing that the response is not a transient oscillation but a metastable reversal of the addressed core coordinate. The radial and out-of-plane components exhibit secondary adjustments, reflecting relaxation of the local polarization environment around the switched core.

The spatial consequence of this same switching event is shown in Fig. 4(a). The relaxed polarization field is compared with the reference texture through difference maps of the in-plane polarization, $\Delta p$, and local topological charge density, $\Delta q$. These maps show that the dominant polarization change and the strongest topological-charge redistribution occur near the addressed core region, while weaker changes extend into neighboring sectors. Neighboring dipoles therefore relax together with the core, and the local topological charge density is redistributed across both inequivalent slices. In the lower slice ($z_1'$), the addressed sector acquires a positive local contribution, whereas a corresponding negative contribution appears in the upper slice ($z_2'$). Additional changes, for example near vortex 5, show that the topological response is collective rather than strictly site-local.

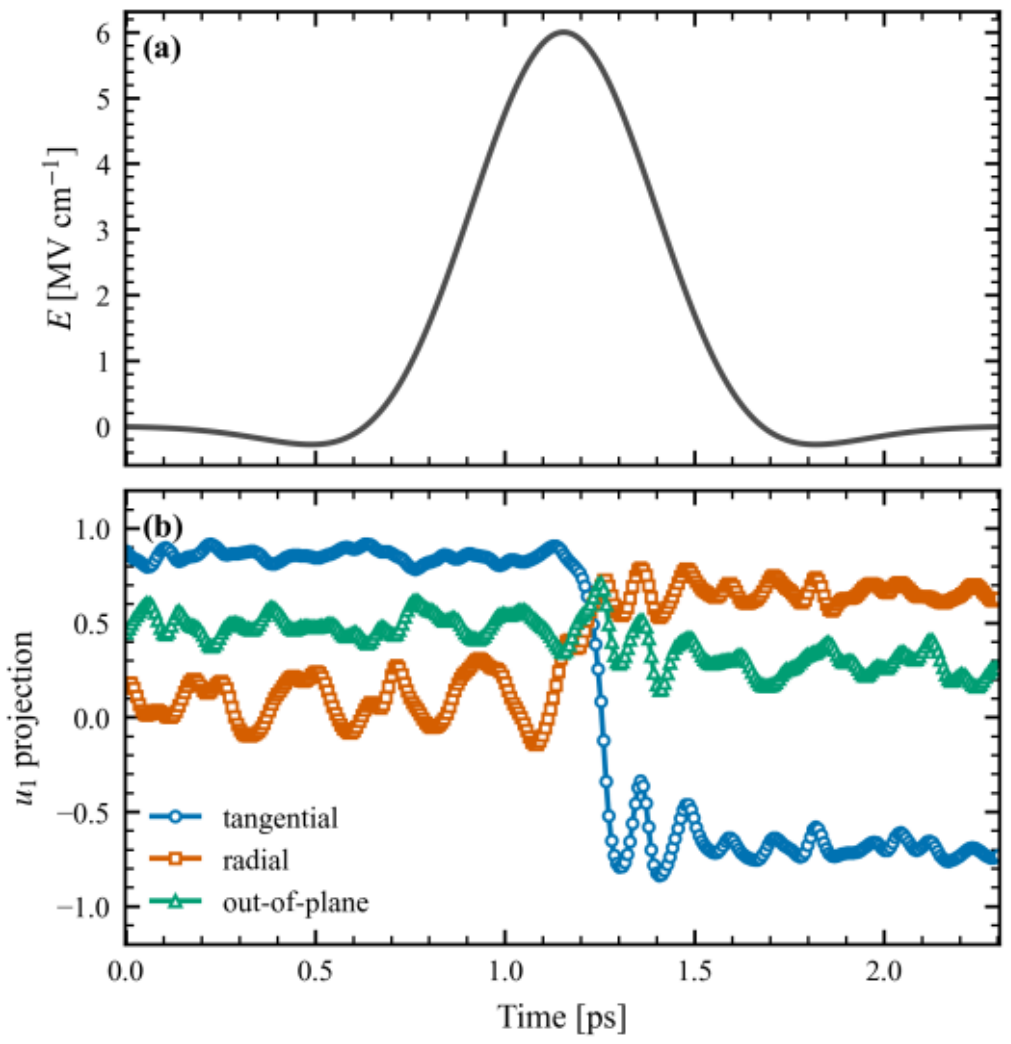

*Figure 3. Time-resolved switching dynamics of the addressed vortex core for pulse mask 100000. (a) Applied local electric-field pulse as a function of time. (b) Time evolution of the local-mode projections of vortex core 1 onto the local tangential, radial, and out-of-plane directions. The tangential component changes sign during the pulse and remains reversed after the field is removed, showing that the excitation drives the core-reversal coordinate across the finite barrier identified in Fig. 2. The radial and out-of-plane components show secondary adjustments associated with relaxation of the local polarization environment.*

This behavior illustrates an important aspect of the switching mechanism. A local reversal of a vortex-core dipole cannot be interpreted as a simple exchange of pre-existing fractional charges between the two slices. Instead, the pulse perturbs one element of a coupled six-vortex scaffold, and the final charge pattern emerges only after collective relaxation of the full nanodomain texture. The redistribution of local charge is constrained by two conditions. First, the slice-integrated topological charge remains quantized. Second, the observed charge transfer between the two inequivalent slices is consistent with mediation by the singular conversion regions separating them. These constraints do not uniquely determine the local charge pattern, but they strongly restrict the allowed rearrangements. In the present system, the local contributions remain organized predominantly in multiples of 1/3, although under excitation partial subdivision into 1/6-like contributions can occur.

A second representative example is the two-core pulse mask 011000, in which vortex cores 2 and 3 are excited simultaneously. The relaxed configuration and corresponding difference maps are shown in Fig. 4(b). As in the single-core case, both targeted vortex cores reverse their in-plane polarization and remain stable after field removal. The corresponding topological response again extends beyond the directly addressed sites. Charge redistribution is observed across both slices of the nanodomain, and the final pattern is determined by collective relaxation of the coupled polarization texture rather than by an independent response of the two targeted vortices. The pulse mask therefore identifies the primary perturbation sites, but not a one-to-one map of final local charge inversion. The same mechanism remains visible for higher-order masks. Fig. 4(c) shows the three-core excitation 101010, in which alternating vortex cores are addressed. The selected cores reverse their polarization direction, while the surrounding dipolar texture relaxes into a new metastable configuration. The associated topological charge density is redistributed over both inequivalent slices, confirming that the final state is governed by the cooperative response of the six-vortex framework. Similar behavior is obtained for other partial pulse masks: local excitation acts as a trigger, but the switched state is selected by relaxation of the full topological nanodomain.

The limiting case is the fully excited mask 111111, shown in Fig. 4(d). Here, the slice-resolved topological sectors are exchanged, yielding a $Q = +4/Q = -2$ sequence. The half of the doubled unit cell that initially hosted the $Q=-2$ antiskyrmionic slice with six $-1/3$ topological quarks now hosts a $Q=+4$ skyrmionic texture with six $+2/3$ quarks, while the opposite half undergoes the inverse conversion. Although the relaxed polarization patterns are not identical to the ground-state slices shown in Fig.

1(a),(b), the topological sectors are fully exchanged. This result shows that the Bloch-point-like singular conversion regions are not merely passive signatures of the reference texture. They therefore appear to provide the local topological channels through which exchange of the slice-resolved charges becomes possible.

All representative switched configurations discussed here remain metastable during the subsequent 1 ns field-free trajectories at 10 K. Their relaxed energies lie close to that of the reference state, indicating that pulse-induced reconfiguration does not require access to a high-energy branch of the nanodomain texture. The persistence of the switched states is therefore attributed to configurational trapping after the pulse has driven the selected vortex-core coordinates across local barriers, followed by relaxation of the surrounding polarization field under the constraints imposed by the coupled six-vortex topology.

These representative examples identify the main switching pathway observed in the simulations. Local ultrafast excitation of selected vortex cores systematically drives the coupled Q=−2/Q=+4 nanodomain into new metastable states, but the resulting reconfiguration is collective rather than strictly local. The following section extends this analysis to the full $2^6$=64 pulse-mask set and examines the resulting programmed states in terms of topological fingerprints, inter-slice charge transfer, and configurational distinguishability.

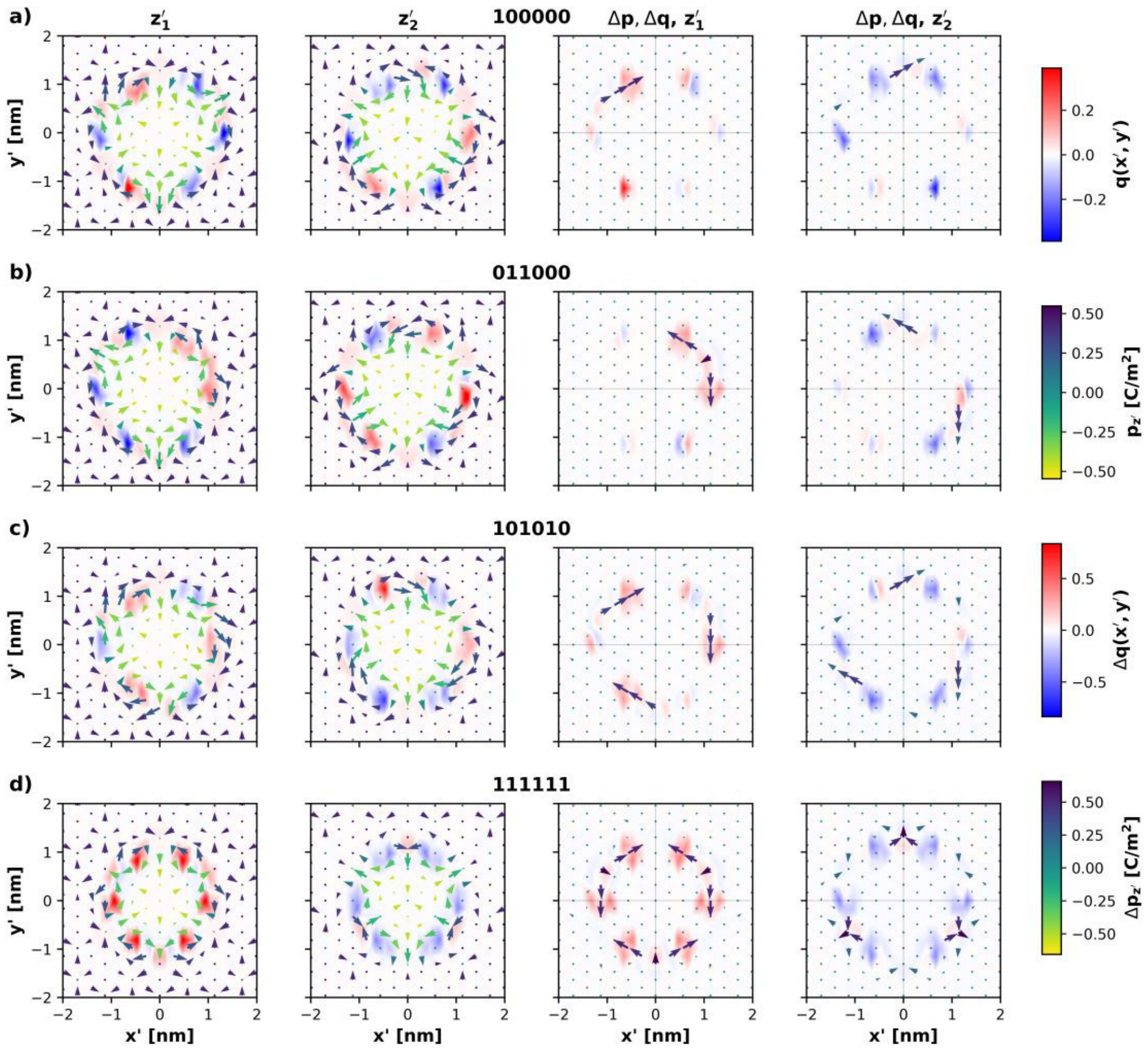

*Figure 4. Representative pulse-induced reconfiguration of the coupled topological nanodomain. For each pulse mask, the first two panels show the relaxed polarization fields and local topological charge densities in the lower $z'_1$ and upper $z'_2$ slices. The last two panels show the corresponding differences relative to the un-pulsed reference texture in Fig. 1. In the difference panels, arrows denote the in-plane polarization change $\boldsymbol{\Delta p} = \boldsymbol{p}_{final} - \boldsymbol{p}_{ref}$, while the color scale denotes the change in local topological charge density $\Delta q(x', y') = q_{final}(x', y') - q_{ref}(x', y')$. Rows show (a) 100000, (b) 011000, (c) 101010, and (d) 111111. The difference maps show that pulse-induced reconfiguration is concentrated near the addressed vortex-core regions but relaxes collectively through the coupled six-vortex scaffold.*

## C. Topological distinguishability of the programmed states

This section extends the analysis from representative switching events to the full set of binary pulse masks. The aim is to determine whether the prescribed local excitation protocol produces distinguishable relaxed topological configurations, rather than to enumerate the complete metastable state space of the continuous dipolar system. Starting from the same relaxed Q = −2/Q = +4 reference texture, all $2^6 = 64$ masks were applied to the six labeled vortex-core columns, followed by 1 ns of field-free evolution at 10 K.

To quantify the local charge redistribution, each of the two inequivalent (111) slices was partitioned into six fixed angular sectors centered on the six vortex positions. The local topological charge density was then integrated within each sector. For the reference pulse protocol, the six-vortex scaffold remains intact after relaxation, with no observed core recombination or annihilation. The same fixed sector decomposition and vortex numbering can therefore be used for all programmed states. The six local charges in the lower slice are denoted $q_1^{(n)}$ with $n = 1, \dots, 6$, and the corresponding charges in the upper slice are denoted $q_2^{(n)}$, using the same vortex numbering as in Fig. 1(c). Each relaxed state is thus represented by a 12-component charge vector, $\boldsymbol{q} = (q_1^{(1)}, \dots, q_1^{(6)}, q_2^{(1)}, \dots q_2^{(6)})$, which serves as a sector-resolved topological fingerprint of the programmed configuration. The full set of fingerprints for all 64 pulse masks, including the 000000-reference state, is shown in Fig. 5(a).

Two general observations follow immediately. First, the pulse protocol is effective across the full set of programmed states: every mask produces a clear redistribution of local topological charge relative to the reference texture. Second, the response is collective rather than strictly local. Excitation of a given subset of vortex cores does not produce a one-to-one modification only of the corresponding sector charges. Instead, the final fingerprint emerges from relaxation of the coupled six-vortex texture as a whole. Even so, systematic trends are evident in Fig. 5(a). In many masks, the sectors associated with the directly excited vortex cores exhibit the strongest changes, showing that the pulse mask still determines the dominant perturbation pattern even though the final state is selected collectively.

The resulting local sector charges are organized predominantly in multiples of 1/3, consistent with the fractional-quark structure of the reference state. In some cases, values close to 1/6-like subdivisions also appear. These values should not be interpreted as exact rational fractions, because the analysis is based on six fixed angular sectors rather than adaptive vortex-centered integration regions. Because the integration sectors are fixed in space, the sector charge depends on how the extended local charge-density feature is distributed within the chosen sector boundaries. The sector values are therefore interpreted as topological fingerprints of the relaxed configurations, rather than as exact rational invariants of individual sectors.

Reproducibility was assessed at the level relevant to the present claim, namely whether the pulse protocol generates distinguishable metastable configurations under a specified excitation protocol, rather than whether every sector-integrated charge remains invariant under changes of the excitation details. For each mask, repeated simulations performed with the same initial state and reference pulse protocol converge to the same switched-state pattern, with only minor variations in the sector-integrated local charges. Small changes in field amplitude or pulse shape can preserve closely related relaxed polarization textures while producing different redistributions of the local topological charge, especially for masks with several excited cores. Reproducibility should therefore be understood here in terms of robust access to metastable reconfigured textures under the specified protocol, rather than as pulse-independent invariance of every individual sector charge.

An important constraint on these local rearrangements is provided by the slice-integrated topological charge. Figure 5(b) shows the total charge of the lower and upper halves of the doubled unit cell for all pulse masks. In every case, the slice-integrated charge remains quantized, within numerical accuracy, to an integer value. The accessible values lie between −2 and +4, i.e. within the range spanned by the two topological sectors of the reference texture. A total slice charge of Q=0 therefore does not imply a topologically trivial state. It can instead arise from cancellation of positive and negative local contributions in the slice integral, while nontrivial local structure remains present. The programmed states therefore preserve integer slice-resolved topology even under strong local excitation.

A second constraint becomes visible when vertically corresponding sectors in the two slices are compared. The local charges $q_1^{(n)}$ and $q_2^{(n)}$ occupy the same in-plane position and are separated along the nanodomain axis by the Bloch-point-like singular conversion region. To characterize the local inter-slice transfer, the sector-wise difference $\Delta q^{(n)} = q_2^{(n)} - q_1^{(n)}$ was evaluated and is shown in Fig. 5(c). The resulting values cluster predominantly near $-1$, 0, or $+1$, within the numerical uncertainty introduced by the fixed-sector decomposition. This behavior indicates that the inter-slice transfer of local topological charge proceeds mainly through integer channels. The singular conversion regions therefore do not merely accompany the switched states structurally. They mediate the integer transfer processes through which the local topological relation between the two slices is reorganized.

These constraints organize the local charge redistribution without fully determining its microscopic pattern. Each switched state is selected by collective relaxation of the coupled six-vortex texture, while the integer slice charges and predominantly integer inter-slice transfer channels remain preserved. The sector-resolved charges therefore provide compact topological fingerprints of the relaxed states generated by the chosen pulse protocol.

Taken together, Figs. 5(b) and 5(c) reveal two robust topological constraints on the programmed state space. First, the integrated charge on each (111) slice remains integer. Second, the difference between vertically corresponding local sectors is restricted predominantly to integer transfer channels. The local charge pattern is thus not free to vary arbitrarily, even though the underlying polarization field is continuous. The observed metastable states emerge from the interplay between local barrier crossing at the driven vortex cores, collective dipolar relaxation of the coupled scaffold, and these discrete topological constraints.

A final question is whether the pulse-addressed states are genuinely distinguishable within this sector-resolved topological representation. To test this, the charge vectors of all states were compared using the $L_\infty$ distance $d_{ij} = \max_k \left| q_k^{(i)} - q_k^{(j)} \right|$, where k runs over the 12 sector charges and $i, j$ label the programmed states. For each state, the minimum distance to its nearest neighbor among the 64 programmed states was then determined. This provides a stringent distinguishability criterion, because it requires that every state differ from every other state by at least one sufficiently large local topological signature. The resulting nearest-neighbor distances are shown in Fig. 5(d). In all cases, the minimum $L_\infty$ distance exceeds 0.75. Thus, no two states share the same sector-resolved topological fingerprint. For every programmed configuration, there exists at least one sector whose charge differs by more than 0.75 from that of the most similar competing state. This separation is large on the scale of the fractional charge units involved and is well above the small deviations from ideal rational values caused by the fixed-sector decomposition.

These distances are not expected to correspond to ideal integer or rational separations, because the fixed-sector analysis is a non-adaptive representation of spatially extended charge-density features. Finite core size and local distortions can distribute charge across neighboring sectors and thereby shift the numerical sector values away from ideal fractional charges. This makes the distinguishability test conservative rather than less meaningful: despite the fixed-sector representation, all 64 pulse-addressed states remain clearly separated in topological space.

The full pulse-mask set therefore supports the central result of the present work: under the chosen low-temperature excitation protocol, the 64 binary masks lead to 64 metastable configurations that are distinguishable by their sector-resolved topological fingerprints. The structural distinguishability of the programmed states in the underlying polarization field is examined in the following section.

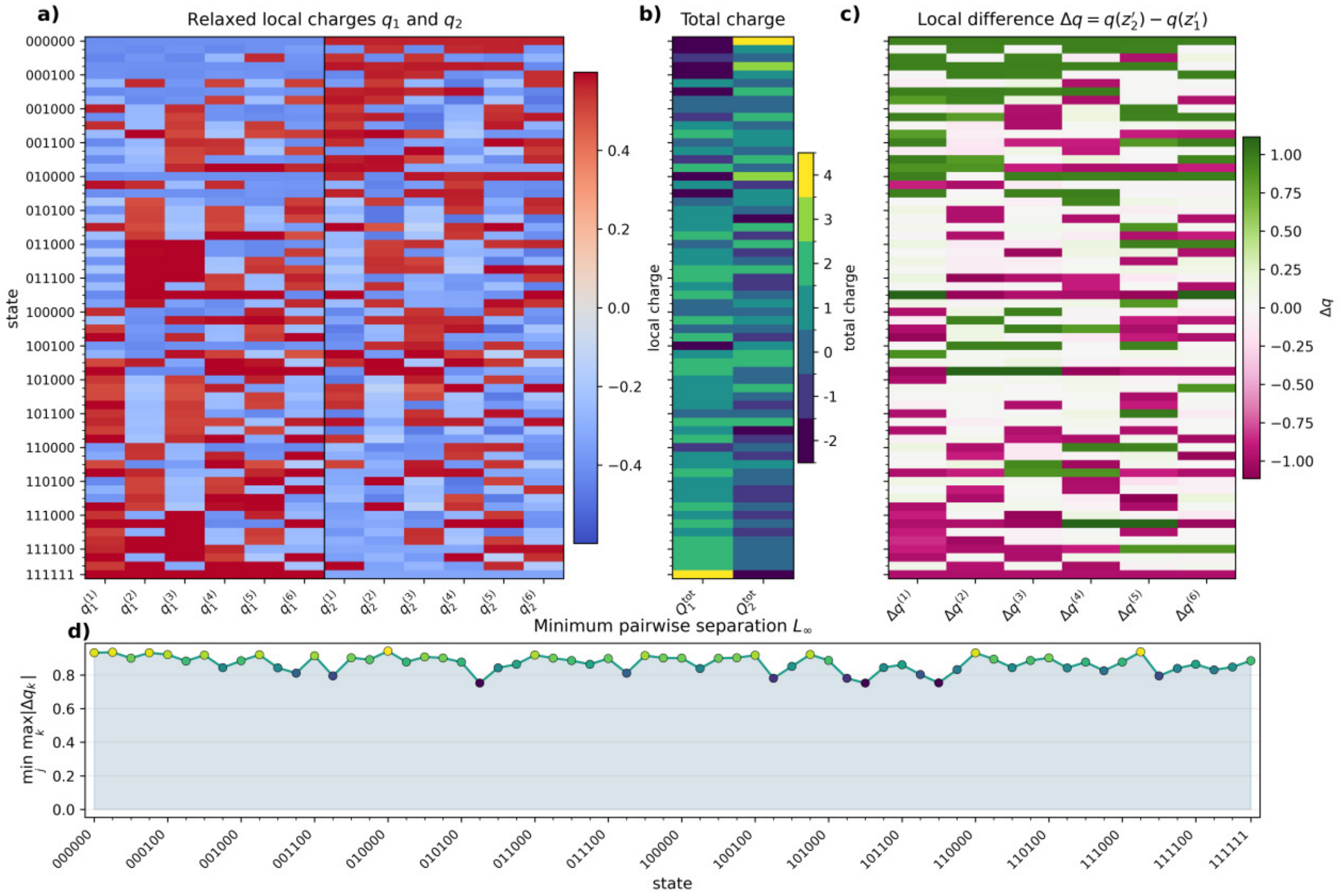

*Figure 5. Topological fingerprints of the full $2^6$=64 pulse-mask set. (a) Sector-integrated local topological charges for all 64 pulse masks, including the 000000-reference state. The lower and upper (111) slices are represented by $q_1^{(n)}$ and $q_2^{(n)}$, respectively, with n=1, ...,6. (b) Slice-integrated topological charges of the lower and upper halves of the doubled unit cell for all pulse masks. (c) Inter-slice sector-wise charge difference, $\Delta q^n = q_2^{(n)} - q_1^{(n)}$, for corresponding vortex positions. (d) Minimum $L_\infty$ distance of each state to its nearest neighbor among the 64 programmed states, demonstrating topological distinguishability of all programmed configurations. All $2^6$=64 pulse masks are shown, with selected axis labels and minor ticks used for visual clarity.*

## D. Polarization-field distinguishability of the programmed states

The sector-resolved topological analysis establishes distinguishability at the level of derived charge fingerprints. It is also useful to compare the underlying polarization fields directly, since the topological charge density is computed from the normalized local-mode field. This section therefore tests whether the pulse-generated configurations remain distinguishable in real-space polarization space, independently of the particular sector integration used in Sec. III.C.

The analysis is based on the time-averaged polarization fields obtained for each programmed state from the final 200 ps of the 1 ns field-free relaxation run. Each state is therefore represented by a three-dimensional vector field $\boldsymbol{p}(\boldsymbol{R}) = (p_x, p_y, p_z)$ defined on the 32×32×32 simulation grid. A direct comparison over the full supercell would be dominated by the nearly unchanged ferroelectric matrix outside the nanodomain. Pairwise comparisons were therefore restricted to a cylindrical region centered on the nanodomain axis, with diameter 6 nm. This region contains the nanodomain and its immediately perturbed environment while excluding the remote matrix. The same mask was used for all states, so the resulting distances provide a fixed-frame measure of structural separation.

Two complementary measures were used to quantify the difference between any pair of programmed states A and B. The first is the pairwise mean angular mismatch. At each masked grid point k, the angle between the two local polarization vectors is evaluated as

$$\theta_k^{AB} = \arccos\left(\frac{\boldsymbol{p}_A^{(k)} \cdot \boldsymbol{p}_B^{(k)}}{\left|\boldsymbol{p}_A^{(k)}\right|\left|\boldsymbol{p}_B^{(k)}\right|}\right), \tag{10}$$

where only points with non-negligible vector norm in both states are included. Averaging $\theta_k^{AB}$ over all valid masked grid points yields the mean angular mismatch $\bar{\theta}_{AB}$. This quantity probes differences in local dipole orientation and is insensitive to the absolute polarization magnitude.

The second measure is the pairwise relative root-mean-square (RMS) vector difference, which probes the full vector-field mismatch including both orientation and magnitude. For a given pair of states, the absolute RMS difference is computed from the local vector difference $\Delta\boldsymbol{p}^{(k)} = \boldsymbol{p}_A^{(k)} - \boldsymbol{p}_B^{(k)}$, and normalized by the mean RMS magnitude of the two original fields. The resulting quantity therefore expresses the structural difference between two polarization textures relative to their characteristic polarization scale. A value of zero corresponds to identical vector fields, whereas larger values indicate progressively stronger structural differences.

The pairwise mean angular mismatch for all state pairs is shown in Fig. 6(a) as a 64×64 matrix. As expected, the diagonal entries vanish and the matrix is symmetric. Regions of enhanced similarity are visible, corresponding to states generated by related pulse masks. In particular, masks that differ only by excitation of a neighboring vortex core often lead to more closely aligned polarization textures than masks that perturb more distinct parts of the nanodomain. This behavior is physically reasonable, since similar excitation patterns act on overlapping regions of the coupled six-vortex structure. Even so, the matrix exhibits a broad spread of values across the full set of programmed states, showing that the 64 states do not collapse into a small number of nearly identical polarization classes.

The analogous all-to-all comparison based on the relative RMS vector difference is shown in Fig. 6(b). The same overall organization of the programmed states is recovered, but the separation becomes more pronounced. This is because the RMS measure captures not only directional changes of the dipoles, but also the accompanying redistribution of polarization magnitude around the switched vortex cores and their surroundings. The programmed states are therefore distinguishable not only in terms of orientation patterns, but also with respect to the full three-dimensional dipolar structure.

To quantify the worst-case separation within the programmed state space, a nearest-neighbor analysis was performed for both metrics. For each of the 64 states, the minimum off-diagonal value of the pairwise mean angular mismatch matrix and the minimum off-diagonal value of the pairwise relative RMS matrix were extracted. These two minima are determined independently and therefore need not correspond to the same competing state. They provide two complementary lower bounds on structural distinguishability. The resulting values are shown in Fig. 6(c). For the mean angular mismatch, all states exhibit minimum values between approximately 2° and 6°. Because these angles are averaged over the full masked nanodomain volume, even values of 2°–6° indicate differences extending over a finite part of the masked region. The relative RMS measure gives a larger separation, with nearest-neighbor values of approximately 0.10–0.25. Because this metric includes both orientation and magnitude changes, it captures the reorganization of the vortex-core columns and their surrounding polarization field more directly than the angular average alone.

These values should be interpreted conservatively. The comparison is carried out within a fixed cylindrical mask and does not adapt to small vortex-core displacements. In addition, the masked region includes not only the strongly reconfigured vortex-core columns, but also the more weakly perturbed surrounding polarization field. Both factors reduce the apparent distance between states. The fact that all 64 states remain clearly separated even under these conditions therefore strengthens the conclusion that the programmed states are structurally well resolved.

This analysis shows that the pulse-generated metastable configurations are not distinguished only by the sector-integrated topological charges. They also occupy distinct positions in polarization-field space under a fixed real-space comparison of the simulated local-mode configurations. A detailed mapping to experimental contrast mechanisms is beyond the scope of this work. The present result instead establishes that the topological fingerprints correspond to structural differences in the underlying simulated polarization textures.

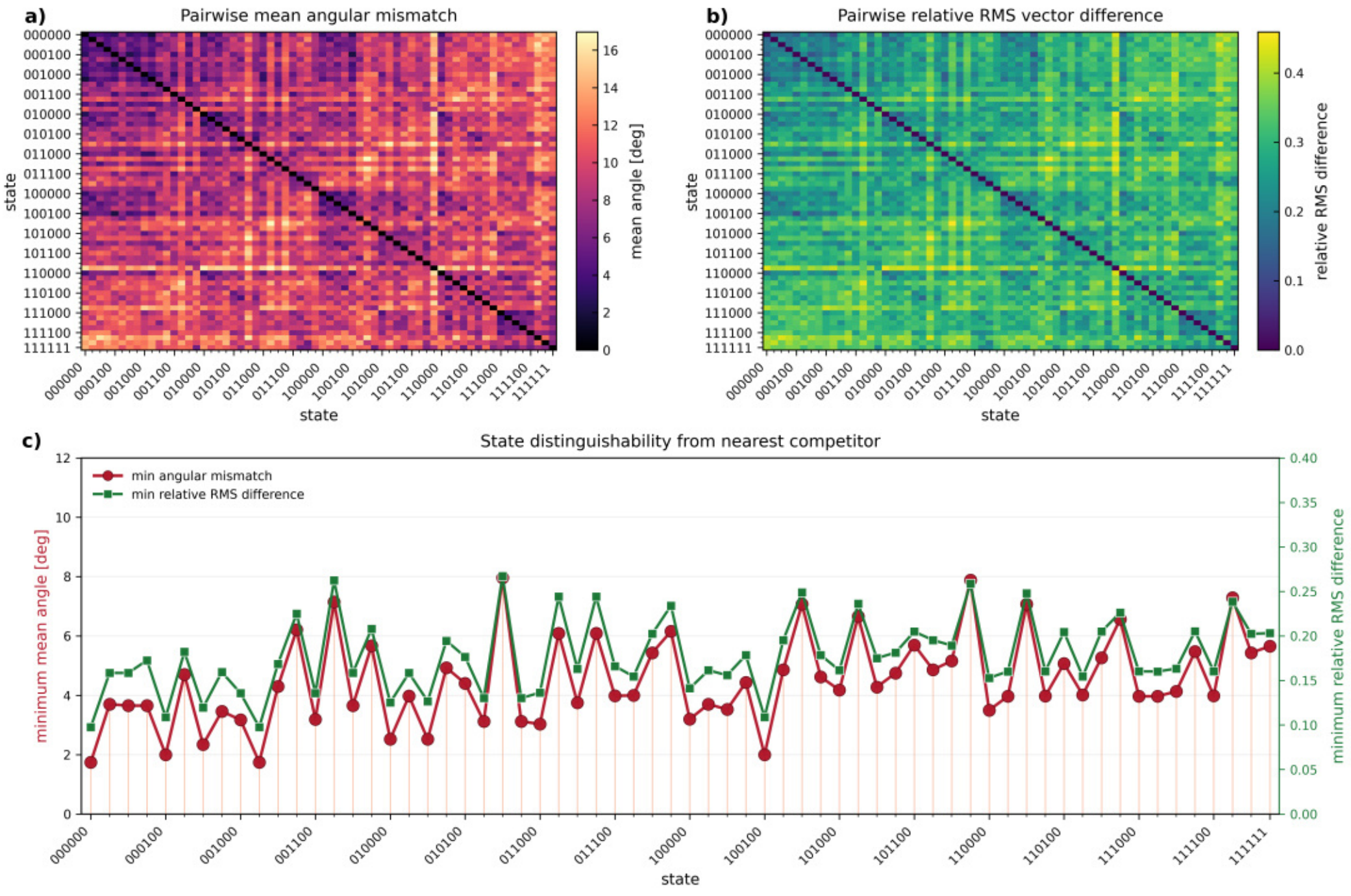

*Figure 6. Polarization-field distinguishability of the programmed states. All comparisons are based on time-averaged polarization fields obtained from the final 200 ps of the field-free relaxation and restricted to a cylindrical mask centered on the nanodomain axis. (a) Pairwise mean angular mismatch between all 64 programmed states. (b) Pairwise relative RMS vector difference between all 64 programmed states. (c) Minimum off-diagonal mean angular mismatch and minimum off-diagonal relative RMS vector difference for each state, quantifying worst-case structural distinguishability in polarization-field space. All $2^6$=64 pulse masks are shown, with selected axis labels and minor ticks used for visual clarity.*

# IV. CONCLUSIONS

The present work examined whether the internal fractional topology of a single ferroelectric nanodomain can be reconfigured by local ultrafast electric excitation. The study focused on ordered 12.5% Zr-substituted BT, where chemical doubling along [111] stabilizes a coupled Q=−2/Q=+4 nanodomain composed of six approximately −1/3 and six approximately +2/3 localized topological charge contributions in two inequivalent (111) slices. This texture provides a model system in which local fractional charge structure, integer slice-resolved topology, and Bloch-point-like conversion regions coexist within the same three-dimensional polarization field.

Effective-Hamiltonian molecular-dynamics simulations show that picosecond local electric-field pulses applied to selected vortex-core columns can drive this coupled texture into distinct field-free metastable configurations. The switching response is collective: the pulse perturbs localized vortex-core motifs, but the final state is selected by relaxation of the full six-vortex scaffold and by coupling between the two inequivalent slices. The fully excited pulse mask exchanges the slice-resolved Q=−2 and Q=+4 sectors, supporting the interpretation that the Bloch-point-like conversion regions provide channels for changes in local inter-slice topological charge. A complementary constrained potential-energy-surface analysis further supports this switching picture by showing that reversal of a single vortex-core column connects locally stable configurations separated by a finite barrier.

Within the binary pulse-mask protocol considered here, the 64 prescribed excitation masks generate 64 relaxed configurations that are distinguishable by sector-resolved topological fingerprints. These configurations retain integer slice-integrated charges and exhibit predominantly integer inter-slice charge differences between corresponding sectors. Direct comparison of the simulated polarization fields further shows that the pulse-generated states are separated in real-space dipolar structure, not only in the derived topological charge representation. The present results therefore establish a low-temperature computational proof of concept that internal fractional polar topology can be locally reconfigured and sampled as a multistate configurational degree of freedom.

The local field profile used in the switching simulations should be viewed as an idealized control protocol. It assumes independently addressable, nanoscale, picosecond field profiles centered on selected vortex-core columns, which is beyond present experimental writing capabilities. The protocol is therefore not intended as a direct experimental implementation, but as a controlled computational perturbation that tests whether fractional polar topology can respond to local electric forcing within the effective-Hamiltonian model. In an experiment, any realization would likely require approximate rather than perfectly prescribed field profiles, or alternative control strategies that exploit the same switchable core coordinates indirectly. The present work should therefore be understood as a computational proof of concept rather than a device-level switching protocol.

The result should be understood in the context of the compact 2.6 nm nanodomain and the cryogenic simulation regime used here. The reference texture remains thermally stable up to about 26 K, while representative pulse-programmed states remain stable within a slightly lower temperature window of approximately 21–26 K, depending on the state. In the representative heating tests, the induced configurations remain identifiable up to temperatures close to nanodomain collapse rather than first relaxing back to the reference $Q=-2/Q=+4$ texture. At 10 K, the switched configurations remain metastable for at least 1 ns of field-free evolution on the simulation timescale. These findings do not imply room-temperature operation or device-level retention, but they do establish a finite low-temperature stability window for pulse-driven reconfiguration of fractional polar topology.

Several routes may be relevant for extending this stability range, including larger nanodomain diameters, epitaxial strain, electrostatic boundary engineering, and controlled compositional modulation. The effect of partial or random B-site substitution also remains an important open question, since disorder may reduce the symmetry and deterministic pulse-to-state relation of the ordered model system. More broadly, the present results suggest that the internal topology of ferroelectric nanodomains can provide a controllable configurational landscape, motivating further studies of stability, disorder tolerance, pulse robustness, and experimentally accessible readout mechanisms. In this sense, ordered 12.5% Zr-substituted BT emerges as a useful model platform for studying pulse-driven reconfiguration of fractional polar topology in the cryogenic regime.

# DATA AVAILABILITY

The data that support the findings of this study are available from the author upon reasonable request.

# ACKNOWLEDGMENTS

The author gratefully acknowledges the financial support under the scope of the COMET program within the K2 Center "Integrated Computational Material, Process and Product Engineering (IC-MPPE)" (Project 886385). This program is supported by the Austrian Federal Ministries for 718 Climate Action, Environment, Energy, Mobility, Innovation, and Technology (BMK) and for Digital and Economic Affairs (BMDW), represented by the Austrian research funding association (FFG), and the federal states of Styria, Upper Austria, and Tyrol.

# Supplemental Material

## Pulse-Driven Reconfiguration of Fractional Polar Topology in Zr-Substituted Barium Titanate

Florian Mayer[1*]

[1]Materials Center Leoben Forschung GmbH, Vordernberger Strasse 12, 8700 Leoben, Austria

*florian.mayer@mcl.at

# I. EFFECTIVE HAMILTONIAN USED FOR SUBSTITUTED BARIUM TITANATE

The molecular-dynamics simulations are based on a first-principles-parameterized effective Hamiltonian in the local-mode framework [1–4]. Each five-atom perovskite unit cell at lattice vector $\boldsymbol{R}$ carries a B-site-centered local soft mode $\boldsymbol{u}(\boldsymbol{R})$, which is proportional to the local dipole moment, and an acoustic displacement variable $\boldsymbol{w}(\boldsymbol{R})$, which describes inhomogeneous strain and local translation degrees of freedom. Homogeneous strain is described by the six Voigt components $\eta_i$, with $i$ = 1,…,6. For the substituted system, each B-site unit cell also carries a discrete species variable $\sigma_R$, which specifies whether the B site is occupied by Ti or Zr.

The parent $BaTiO_3$ Hamiltonian follows the standard Nishimatsu-type [1,2] effective-Hamiltonian form. It can be written schematically as

$$\begin{aligned}H_{BTO}^{eff} = \frac{M_{dipole}^{*}}{2}\sum_{R,\alpha}\dot{u}_{\alpha}^{2}(\boldsymbol{R}) \\ + \frac{M_{acoustic}^{*}}{2}\sum_{R,\alpha}\dot{w}_{\alpha}^{2}(\boldsymbol{R}) + V^{self}(\{\boldsymbol{u}\}) + V^{dpl}(\{\boldsymbol{u}\}) + V^{short}(\{\boldsymbol{u}\}) \\ + V^{elas,homo}(\eta_1,\dots,\eta_6) + V^{elas,inho}(\{\boldsymbol{w}\}) + V^{coupl,homo}(\{\boldsymbol{u}\},\eta_1,\dots,\eta_6) \\ + V^{coupl,inho}(\{\boldsymbol{u}\},\{\boldsymbol{w}\})\end{aligned} \tag{S1}$$

Here $M_{dipole}^{*}$ and $M_{acoustic}^{*}$ are the effective masses associated with the local-mode and acoustic variables, respectively. The terms $V^{self}$, $V^{dpl}$, and $V^{short}$ denote the local-mode self-energy, the long-range dipole-dipole interaction, and the short-range interaction between local modes. The terms $V^{elas,homo}$ and $V^{elas,inho}$ describe the homogeneous and inhomogeneous elastic energies, while $V^{coupl,homo}$ and $V^{coupl,inho}$ describe the coupling of the local modes to homogeneous and inhomogeneous strain. The explicit polynomial forms and parameter values are those of the $BaTiO_3$ effective-Hamiltonian parameterization described in Refs. [1–4]. Equation (S1) denotes the field-free parent Hamiltonian. The electric-field coupling used in the pulse simulations is introduced separately in Eq. (S6), where the standard homogeneous-field term is generalized to a spatially and temporally varying field with species-dependent Born effective charges.

B-site substitution is introduced as a perturbation of the parent $BaTiO_3$ Hamiltonian [5]. The total substituted Hamiltonian is written as

$$H_{total}^{eff}(\{\boldsymbol{u}\},\{\boldsymbol{w}\},\eta_i,\{\sigma\}) = H_{BTO}^{eff}(\{\boldsymbol{u}\},\{\boldsymbol{w}\},\eta_i) + H^{peturb.}(\{\boldsymbol{u}\},\{\boldsymbol{w}\},\{\sigma\}) \tag{S2}$$

with

$$H^{peturb.} = \Delta T(\{\boldsymbol{u}\},\{\boldsymbol{w}\},\{\sigma\}) + \Delta V^{self}(\{\boldsymbol{u}\},\{\sigma\}) + \Delta V^{dpl}(\{\boldsymbol{u}\},\{\sigma\}) + V^{aux}(\{\boldsymbol{u}\},\{\boldsymbol{w}\},\{\sigma\}) \tag{S3}$$

The first term, $\Delta T$, accounts for species-dependent effective masses. The term $\Delta V^{self}$ modifies the local-mode self-energy on substituted B sites. The term $\Delta V^{dpl}$ accounts for the species dependence of the Born effective charges entering the long-range dipole-dipole interaction. The auxiliary term $V^{aux}$ captures local chemistry-induced changes of interactions beyond the parent $BaTiO_3$ Hamiltonian. The detailed fitting procedure and the explicit forms of the species-dependent corrections are given in Ref. [5].

The local-chemistry correction is included to first order through the auxiliary term

$$V^{aux}(\{\boldsymbol{u}\},\{\boldsymbol{w}\},\sigma) = \sum_R \sum_T Q_{\boldsymbol{T},\boldsymbol{R}}(\sigma_{\boldsymbol{T}})\boldsymbol{e}_{\boldsymbol{T},\boldsymbol{R}} \cdot \boldsymbol{u}(\boldsymbol{R}) + \sum_R \sum_T S_{\boldsymbol{T},\boldsymbol{R}}(\sigma_{\boldsymbol{T}})\boldsymbol{f}_{\boldsymbol{T},\boldsymbol{R}} \cdot \boldsymbol{w}(\boldsymbol{R})$$

(S4)

Here $\boldsymbol{T}$ runs over neighboring cells up to the third-nearest-neighbor shell. The unit vector $\boldsymbol{e}_{\boldsymbol{T},\boldsymbol{R}}$ points from site $\boldsymbol{T}$ to the center of the local mode $\boldsymbol{u}(\boldsymbol{R})$, and $\boldsymbol{f}_{\boldsymbol{T},\boldsymbol{R}}$ is defined analogously for the acoustic variable $\boldsymbol{w}(\boldsymbol{R})$. The coefficients $Q_{\boldsymbol{T},\boldsymbol{R}}$ and $S_{\boldsymbol{T},\boldsymbol{R}}$ depend on the species occupying site $\boldsymbol{T}$ and are obtained from first-principles fitting as described in Ref. [5].

For Zr-substituted $BaTiO_3$, the dopant-induced change of the equilibrium lattice parameter is incorporated through a composition-dependent hydrostatic pressure correction. Following Ref. [5], the pressure correction is taken as

$$P(x) = -19.5x\ [GPa]$$

(S5)

where $x$ is the Zr concentration. The homogeneous strain variables are therefore relaxed self-consistently under this imposed composition-dependent pressure rather than under strictly zero external pressure.

The coupling to an external electric field is introduced through the local dipole associated with each local mode. For a spatially and temporally varying applied field $\boldsymbol{E}(\boldsymbol{R},t)$, the field contribution to the Hamiltonian is

$$H_E(t) = -\sum_{\boldsymbol{R}} Z^*(\sigma_{\boldsymbol{R}})\boldsymbol{E}(\boldsymbol{R},t) \cdot \boldsymbol{u}(\boldsymbol{R})$$

(S6)

Here $Z^*(\sigma_{\boldsymbol{R}})$ is the Born effective charge of the local mode associated with the B-site species $\sigma_{\boldsymbol{R}}$. Thus Ti-centered and Zr-centered unit cells couple to the applied electric field with different effective charges. The same species-dependent Born effective charges are used in the long-range dipole-dipole interaction of the substituted effective Hamiltonian.

The corresponding electric-field force acting on the local mode is

$$\boldsymbol{F}_E(\boldsymbol{R},t) = -\frac{\partial H_E}{\partial \boldsymbol{u}(\boldsymbol{R})} = Z^*(\sigma_{\boldsymbol{R}})\boldsymbol{E}(\boldsymbol{R},t)$$

(S7)

The species-dependent Born effective charges used in the substituted Hamiltonian were obtained from cubic $BaTiO_3$ and $BaZrO_3$ reference calculations within density-functional theory, consistently with the construction of the local-mode parameterization. They are treated as fixed B-site-species parameters, so that Ti- and Zr-centered local modes couple to the applied field with different strengths. This approximation is standard for effective-Hamiltonian simulations. Environment-dependent corrections to $Z^*$ are not included and would primarily renormalize the quantitative field scale.

The local polarization associated with a unit cell can be written as

$$\boldsymbol{p}(\boldsymbol{R}) = \frac{Z^*(\sigma_{\boldsymbol{R}})}{\Omega_0}\boldsymbol{u}(\boldsymbol{R})$$

(S8)

where $\Omega_0$ is the unit-cell volume. For the topological analysis in the main text, the normalized order-parameter field is constructed as

$$\boldsymbol{n}(\boldsymbol{R}) = \frac{\boldsymbol{u}(\boldsymbol{R})}{|\boldsymbol{u}(\boldsymbol{R})|}$$

(S9)

or equivalently from the local polarization direction. Since the topological charge depends on the direction of the local polar order parameter, the normalization removes the amplitude and retains the mapping of the local polar texture to the unit sphere.

# II. ENERGETIC LANDSCAPE OF SINGLE-VORTEX-CORE REVERSAL

To assess whether local switching of a vortex core is energetically plausible, the constrained potential-energy surface associated with reversal of a single vortex-core column was examined explicitly. The purpose of this analysis is not to reproduce the full nonequilibrium pulse-driven dynamics discussed in the main text, but to determine whether the ordered 12.5% Zr-substituted $BaTiO_3$ nanodomain contains more than one locally stable vortex-core configuration and whether these configurations are separated by a finite barrier. The calculations were performed within the effective-Hamiltonian framework at 1 K using the same 32×32×32 supercell employed in the main study. As initial state, the relaxed cylindrical nanodomain containing the coupled Q=−2/Q=+4 texture was used. Vortex core 1, labeled according to Fig. 1(c) of the main text, was selected as the constrained degree of freedom. More precisely, the local-mode vector associated with this vortex core was constrained along the full nanodomain axis z′, such that the same local-mode vector was imposed for the entire vortex-core column. The resulting energy landscape therefore corresponds to a columnar switching coordinate embedded in the full topological nanodomain. To construct the constrained potential-energy surface, a three-dimensional grid of local-mode amplitudes was sampled. The three Cartesian components of the imposed local mode were varied from −0.17 to +0.17 Å using 21 grid points per direction. For each grid point, the selected vortex-core column was held fixed while all remaining degrees of freedom were relaxed. The resulting constrained energy landscape is therefore $E(u_x, u_y, u_z)$, where $(u_x, u_y, u_z)$ denotes the imposed local-mode vector of the selected vortex-core column.

The resulting potential-energy surface exhibits two distinct local minima. One corresponds to the reference vortex-core orientation of the relaxed ground-state nanodomain, i.e. the configuration shown in Fig. 1 of the main text. The second corresponds to a metastable configuration associated with reversal of the vortex-core state, although the relaxed path through the constrained landscape is not required to follow a simple antiparallel displacement of the initial tangential local-mode vector. Instead, the minimum-to-minimum connection reflects the full relaxed response of the constrained core column embedded in the surrounding nanodomain. The existence of this second minimum shows that the vortex-core column is not described by a purely harmonic local landscape around the reference state, but instead possesses at least two locally stable configurations within the constrained nanodomain environment.

To characterize the barrier separating these minima, both minima were first identified in the full three-dimensional constrained energy landscape $E(u_x, u_y, u_z)$. A discrete minimum-barrier path search was then performed on the sampled grid by minimizing the maximum energy encountered along connected grid paths between the two minima, denoted M1 and M2. This search showed that, within the resolution of the sampled grid, the lowest-barrier connected route follows the direct M1→M2 direction. Accordingly, Fig. S1(a) shows a two-dimensional slice of the potential-energy surface passing through both minima and containing this path, while Fig. S1(b) shows the corresponding energy profile along the same route. Because M1 and M2 are not isoenergetic, the barrier depends on the direction of traversal. Along the path shown in Fig. S1(b), the profile rises by approximately 1.8 eV from the lower-energy state M1 to the highest point on the connecting path, whereas the barrier from the metastable state M2 toward M1 is approximately 0.4 eV for the full 32×32×32 simulation cell. The latter quantity is the relevant local barrier for escape from the metastable vortex-core configuration identified here. Within the present constrained coordinate, this establishes that reversal of the selected vortex-core column involves a finite energetic threshold rather than a barrierless relaxation back to the reference state.

This analysis provides a direct energetic basis for the switching picture developed in the main manuscript. The local pulse does not act on a featureless harmonic degree of freedom, but on a pre-existing multi-well local coordinate: the selected vortex-core column already possesses two locally stable configurations separated by a finite barrier. The role of the external pulse is therefore to drive the system across this barrier, after which the surrounding polarization field can relax into a corresponding metastable topological state. In this sense, the constrained potential-energy surface supports the interpretation that the programmed states reported in the main text are rooted in an underlying multi-well energy landscape of the coupled nanodomain texture rather than in accidental transient distortions.

Several limitations of this mapping should be noted. First, the analysis was carried out for a single constrained vortex-core column only, i.e. for the coordinate $E(u_{x,v_1}, u_{y,v_1}, u_{z,v_1})$. The landscape will generally change when multiple vortex cores are driven simultaneously, because the six-vortex scaffold is strongly coupled through electrostatic, short-range, and elastic interactions. Second, the surface shown here is a constrained relaxation landscape sampled on a finite three-dimensional grid. Although the path shown in Fig. S1(b) was identified as the lowest-barrier connected route within this sampled grid, it should not be interpreted as a formally exact transition state or as a continuum minimum-energy path of the type obtained from a dedicated saddle-point or nudged-elastic-band calculation. The reported ~0.4 eV value should therefore be understood as a grid-resolved barrier for escape from the metastable state M2 toward M1 along the chosen collective coordinate. Third, the calculation was performed at 1 K in order to suppress thermal broadening and resolve the intrinsic energetic structure as clearly as possible. It therefore complements, rather than replaces, the low-temperature pulse-switching simulations discussed in the main text. Despite these limitations, the conclusion is clear. The ordered 12.5% Zr-substituted $BaTiO_3$ nanodomain contains at least two locally stable configurations for a given vortex-core column. The metastable configuration M2 is separated from the lower-energy state M1 by a finite barrier of about 0.4 eV for the full 32x32x32 simulation cell, providing an energetic basis for pulse-driven switching of the vortex-core state. The existence of this local bi-stability supports the central mechanism proposed in the main manuscript: vortex cores act as switchable local motifs embedded in a collectively constrained fractional topological texture.

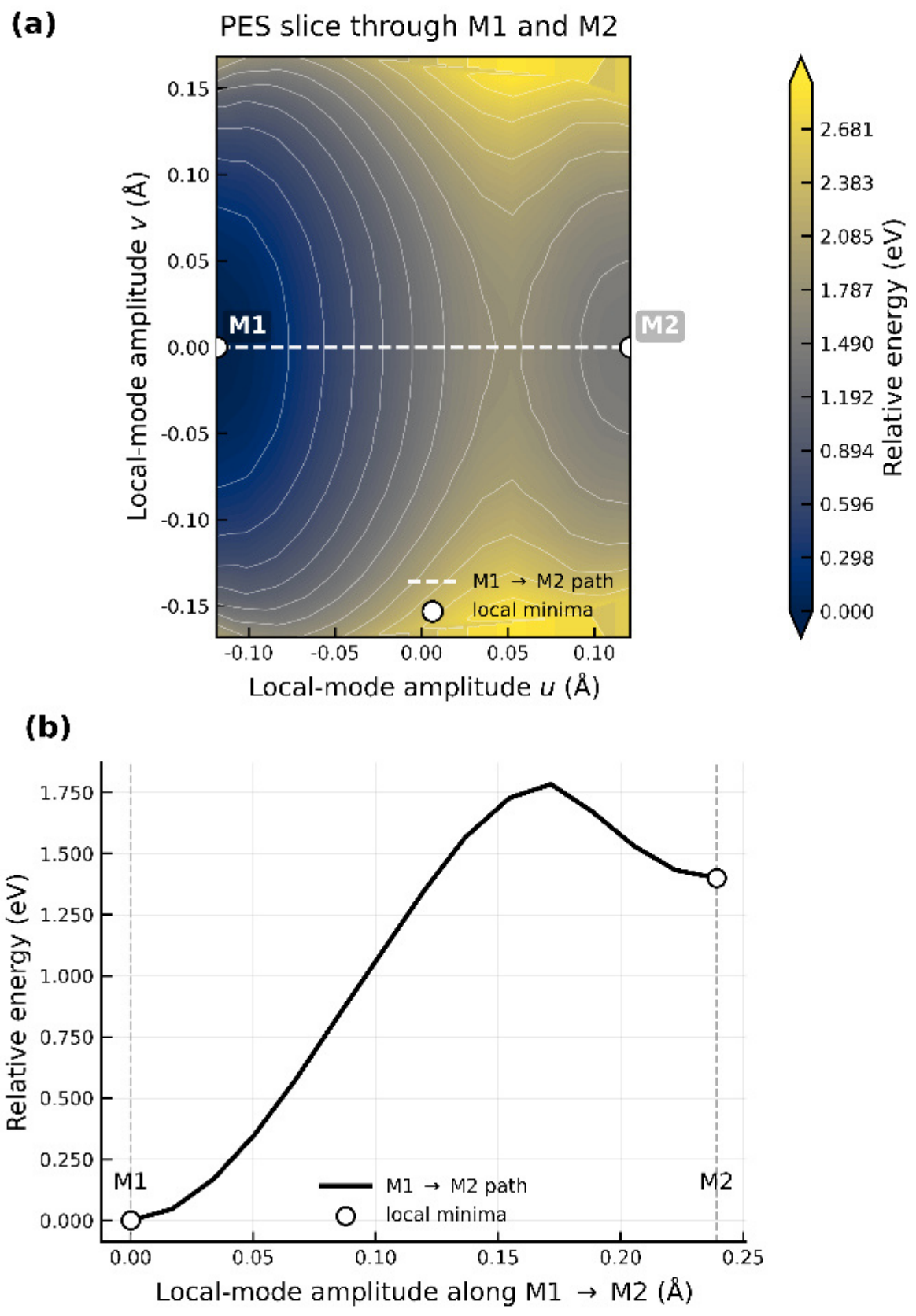

*Figure S1. Constrained energy landscape for switching of a single vortex-core column. (a) Two-dimensional slice of the constrained potential-energy surface $E(u_x,u_y,u_z)$ passing through the two local minima M1 and M2. The plotted coordinates are local-mode amplitudes in Å, and the color scale gives the relative energy in eV for the full 32×32×32 simulation cell. The dashed line denotes the connecting M1→M2 direction. (b) Relative-energy profile along the same path. A discrete minimum-barrier search through the full sampled three-dimensional grid identified this path as the lowest-barrier connected route between the two minima within the grid resolution. The barrier from the metastable state M2 toward M1 is approx. 0.4 eV.*

# III. ALTERNATIVE EXCITATION SCHEMES AND PULSE-SHAPE ROBUSTNESS

The main text focuses on the excitation protocol in which the local writing fields associated with the selected vortex cores are applied simultaneously by superposition, while the nominally unselected vortex cores are not driven directly. To assess whether the reported multistate reconfiguration depends critically on this specific choice, two additional excitation schemes were also examined: sequential excitation of the selected vortex cores and simultaneous excitation combined with stabilizing fields at the nominally unexcited vortex-core positions.

In the sequential protocol, the selected vortex cores are driven one after another rather than simultaneously. In the stabilization protocol, the selected vortex cores are driven simultaneously as in the main text, while auxiliary local fields are applied at the nominally unexcited vortex cores in order to hold them closer to their reference orientations during the pulse-driven evolution. In both cases, the qualitative behavior remains the same as for the protocol used in the main text. All $2^6$=64 pulse masks again generate metastable reconfigured states, and the resulting programmed textures remain distinguishable at the level of their sector-resolved topological fingerprints. The differences relative to the main-text protocol are quantitative rather than qualitative. In particular, individual sector-integrated topological charges can shift because the system is driven through a different dynamical pathway before relaxing into a nearby metastable minimum. This behavior is expected for a continuous coupled dipolar texture in which the local pulse acts on switchable vortex-core coordinates, whereas the final state is selected only after collective relaxation of the full six-vortex scaffold. The main conclusion of the manuscript therefore does not depend on a unique or fine-tuned excitation scheme.

The robustness of the switching behavior with respect to the temporal pulse form was also examined. For the protocol comparison above, the same pulse shape as in the main text was used throughout in order to isolate the effect of the spatial excitation scheme. Additional tests with modified pulse amplitudes and durations show that switching remains robust over a finite parameter range. If the local field amplitude is chosen too small, some targeted states are no longer reached reliably and the system can relax back to the reference texture. At the opposite extreme, particularly under simultaneous excitation of several vortex cores, sufficiently large amplitudes can induce a reorientation strong enough that the nanodomain loses its integrity and collapses. Thus, the useful field window depends on both the pulse mask and the temporal pulse parameters.

The role of pulse duration was tested explicitly for the single-core mask 100000. For the reference temporal waveform used in the main text, the approximate threshold for persistent reversal of the addressed vortex-core coordinate is $E_{sw}(\Delta_0) \simeq 5.5$ MV cm$^{-1}$, so the value $E_0 = 6$ MV cm$^{-1}$ used in the main pulse-mask calculations lies slightly above this single-core switching threshold. The full temporal waveform was then uniformly rescaled while preserving its shape. Starting from the reference waveform with width $\Delta_0 = 0.45$ ps and frequency $\omega_0$=3 ps$^{-1}$, a scaling factor $s$ was applied such that $\Delta = s\Delta_0$ and $\omega = \omega_0/s$. The peak field was varied to determine the approximate threshold for persistent core reversal after field removal. Compressing the waveform by a factor of two ($s$ = 1/2) increases the switching threshold to approximately $1.15E_{sw}(\Delta_0)$, corresponding to about 6.3 MV cm$^{-1}$. Stretching the waveform by a factor of two ($s$ = 2) reduces the threshold to approximately $0.9E_{sw}(\Delta_0)$, corresponding to about 5.0 MV cm$^{-1}$. Thus, the response is not determined by the instantaneous peak field alone, but by the finite-time dynamical action of the pulse on the vortex-core coordinate. This behavior is consistent with earlier ultrafast effective-Hamiltonian simulations showing that field-driven ferroelectric transformations can depend on the temporal pulse history [6], although the present data do not imply a unique scaling with a single integrated field quantity.

It should be emphasized that the nominal field amplitudes used here appear large when compared directly with experimental switching fields. Two points are relevant in this context. First, effective-Hamiltonian molecular-dynamics simulations are known to overestimate field scales relative to experiment, often by approximately one order of magnitude. Second, the present pulses are both highly localized and ultrashort, so comparatively large nominal amplitudes are required in order to produce a measurable dynamical response within the accessible simulation window. The fields reported here should therefore be interpreted primarily as simulation-level control parameters for accessing the metastable reconfiguration pathways of the nanodomain, rather than as direct quantitative predictions of experimental threshold fields.

# IV. THERMAL STABILITY OF THE REFERENCE AND PROGRAMMED NANODOMAIN STATES

In order to assess the thermal robustness of the coupled Q=−2/Q=+4 nanodomain texture, additional finite-temperature simulations were carried out for both the unperturbed reference state and a representative set of pulse-programmed excited states. The purpose of this analysis is not to provide an exhaustive thermal phase diagram of the full 64-state manifold, but to establish the finite temperature window over which the compact nanodomain geometry used in the main text remains stable and to determine whether the pulse-induced states persist over a comparable low-temperature range.

For the reference texture, heating runs were performed for cylindrical nanodomains of different diameters. Starting from the relaxed low-temperature Q=−2/Q=+4 nanodomain, the temperature was increased in steps of 1 K. At each temperature step, the system was evolved for 200 ps before the next increment was applied. The collapse temperature was defined as the first temperature at which the nanodomain disappeared and the system relaxed into a single-domain configuration. Within this protocol, the collapse temperature depends strongly on nanodomain diameter. The smallest stable nanodomains occur only slightly above 2.2 nm and remain stable up to about 11 K, whereas larger nanodomains exhibit substantially higher collapse temperatures. For the diameters examined here, the clean reference Q=−2/Q=+4 texture remains stable up to approximately 26 K for a 2.6 nm nanodomain, 61 K for a 3.2 nm nanodomain, and 66 K for a 3.8 nm nanodomain. These results show that the coupled reference topology is not restricted to a narrow size range and that nanodomain diameter provides an effective parameter for tuning thermal robustness.

The pulse-programmed excited states were examined using the same heating protocol. Fully relaxed switched configurations obtained at 10 K were used as initial states, and the temperature was increased in steps of 1 K with 200 ps evolution at each step until collapse. Because the full $2^6$=64 pulse-addressed set is large, representative states were selected rather than testing every mask individually. The tested set includes both weakly perturbed and strongly reconfigured cases, such as 100000, 100001, 111000, 101010, and 111111, together with additional masks spanning different switching patterns. Collapse was defined, as for the reference state, by loss of the nanodomain and relaxation into a single-domain configuration.

The representative pulse-programmed states collapse between approximately 21 and 26 K, depending on the state, slightly below the collapse temperature of the clean 2.6 nm reference nanodomain. Importantly, the induced metastable configurations do not first relax back into the reference Q=−2/Q=+4 texture during heating. Instead, they remain identifiable up to temperatures close to nanodomain collapse. This shows that the switched configurations are not merely short-lived post-pulse distortions, but metastable states with finite thermal stability in the cryogenic regime.

These calculations therefore identify a finite low-temperature stability window for programmable fractional polar topology in the compact 2.6 nm nanodomain. The main pulse-switching simulations were carried out at 10 K, well below the collapse temperature of both the reference texture and the tested excited states. The larger collapse temperatures found for 3.2 and 3.8 nm nanodomains further indicate that thermal stability is not a fixed limitation of the coupled Q=−2/Q=+4 topology itself, but depends sensitively on nanodomain geometry. This supports the view advanced in the main text that higher-temperature operation should, in principle, be approachable through nanodomain-size optimization and further materials-design strategies such as strain or electrostatic boundary engineering.